\documentclass[11pt]{article}
\pdfoutput=1

\usepackage{jheppub}
\usepackage{bigints}

\usepackage{bbm}

\usepackage{bbold}

\usepackage{tabularx}

\usepackage{tikz}

\usepackage{mathabx}

\usepackage{ae}
\usepackage[T1]{fontenc}
\usepackage{url}

\usepackage{tabularx}

\usepackage{graphicx}
\usepackage{dsfont}
\usepackage{setspace}
\usepackage{amsfonts,amsmath,amsthm,amssymb,amsbsy}
\usepackage{longtable}
\usepackage[latin1]{inputenc}
\usepackage{array}
\usepackage{color}
\usepackage{needspace}
\usepackage[numbers]{natbib}
\usepackage{colortbl}

\usepackage{enumitem}

\usepackage{fancyhdr}

\title{The Loop Momentum Amplituhedron}

\author[]{Livia Ferro}\emailAdd{l.ferro@herts.ac.uk}
\author[]{and Tomasz \L ukowski,}\emailAdd{t.lukowski@herts.ac.uk}

\affiliation[]{Department of Physics, Astronomy and Mathematics, \\ University of Hertfordshire, \\  Hatfield, Hertfordshire, AL10 9AB, United Kingdom}

\abstract{In this paper we focus on scattering amplitudes in maximally supersymmetric Yang-Mills theory and define a long sought-after geometry, the loop momentum amplituhedron, which we conjecture to encode tree and (the integrands of) loop amplitudes in spinor helicity variables.  Motivated by the structure of amplitude singularities, we define an extended positive space, which enhances the Grassmannian space featuring at tree level, and a map which associates to each of its points  tree-level kinematic variables and loop momenta. The image of this map is the loop momentum amplituhedron. 
Importantly, our formulation provides a global definition of the loop momenta. We conjecture that for all multiplicities and helicity sectors, there exists a canonical logarithmic differential form defined on this space, and provide its explicit form in a few examples.
}

\begin{document}
\begin{flushright}
{}
\end{flushright}
\maketitle


\section{Introduction}

In recent years we have seen tremendous advances in new geometric formulations of observables in  Quantum Field Theories, known nowadays under the name of positive geometries \cite{Arkani-Hamed:2017tmz}. These are defined recursively as regions with boundaries of all codimensions, where each boundary is again a positive geometry. Importantly, they are equipped with a unique differential form with logarithmic singularities along all boundaries -- the canonical form -- which, for physically relevant positive geometries, is a physical quantity.
A striking feature of these geometries is that from their definition, by imposing only positivity constraints on some external data, they encode physical properties, such as locality and unitarity, in their boundary structure. 

In this paper we focus on positive geometries for scattering amplitudes in $\mathcal{N}=4$ super Yang-Mills (sYM) theory, i.e. amplituhedra -- see \cite{Ferro:2020ygk,Herrmann:2022nkh} for extensive reviews. Two geometries have been defined in this theory: the amplituhedron \cite{Arkani-Hamed:2013jha} and the momentum amplituhedron \cite{Damgaard:2019ztj}. The amplituhedron, which has been the prime example of a positive geometry, is defined in momentum twistor space and is relevant for tree- and loop-level expectation values of Wilson loops, i.e. scattering amplitudes with the maximally-helicity-violating (MHV) part factored out. Importantly, since ordering is crucial in the definition of momentum twistors, the amplituhedron encodes only the planar sector of  $\mathcal{N}=4$ sYM. The momentum amplituhedron is instead formulated in the non-chiral spinor helicity space and therefore provides a natural language to extend the positive geometry construction to the non-planar sector. However, until now, it was defined only for amplitudes at tree level. The natural question arises whether there exists a positive geometry formulated in the non-chiral spinor helicity space which also produces the amplitude integrands at loop level. In this paper we provide an affirmative answer to this question and formulate the long sought-after geometry for loop amplitudes in spinor helicity space -- the {\it loop momentum amplituhedron}. 

Finding the loop momentum amplituhedron has been a long-standing and important goal since the inception of the tree-level geometry. One of the main obstacles that had been hindering this construction was to find a proper, global definition of the off-shell loop momentum in the spinor helicity space. Indeed, while in the dual space, where momentum twistors are defined, the loop momentum is uniquely determined up to a global shift, in the Feynman approach in momentum space we can redefine it independently for each Feynman diagram. The final answer, after performing Feynman integrals, does not depend on these redefinitions, however, the integrand itself changes significantly. In particular, this leads to introducing unphysical singularities that should not arise in the geometric approach. 
In this paper we resolve this problem by providing a global definition of the loop momenta that serve as parameters in the map defining the loop momentum amplituhedron. To construct this map, we emphasize an important fact about the singularity structures of scattering amplitudes and expectation values of Wilson loops. While at tree level their singularities differ, at loop level there is a one-to-one correspondence between the singularities of integrands for these quantities. Then, since amplituhedra encode physical singularities in the structure of their boundaries, this statement is valid for the boundaries of the geometries: while at tree level the amplituhedron and momentum amplituhedron boundaries are different, at loop level they can be mapped to each other in a simple way. Led by this observation, we draw inspiration from the loop amplituhedron and find its counterpart in spinor helicity variables. For this purpose, we use the relation between momentum twistors and spinor helicity variables, which first appeared in the Grassmannian formulations of scattering amplitudes in these two spaces \cite{Arkani-Hamed:2009nll}.
This allows us to define the loop momentum amplituhedron. By enhancing the Grassmannian space $G_+(k,n)$ present at tree level with $L$ two-planes and requiring additional positivity constraints, we find that the $L$-loop momentum amplituhedron is the image of a map which associates to every point of this extended positive space the tree-level variables $(\lambda,\tilde\lambda)$ and the loop momenta $\ell_p$ for $p=1,\ldots ,L$.

This paper is organized as follows. In section \ref{sec:amplituhedra} we review the basic facts about amplituhedra. In particular, we start by recalling, and afterwards refining, the definition of the momentum amplituhedron at tree level. After a review of the amplituhedron, we present the motivation for our definition of momentum amplituhedron at loop level. Section \ref{sec:loopmomamp} is the main part of the paper and contains the definition of the loop momentum amplituhedron. We then present a few examples in section \ref{sec:examples}, before closing with conclusions and outlook. Appendix \ref{app:vars}  collects the definitions of all variables used throughout the paper.


\section{Amplituhedra}
\label{sec:amplituhedra}

In this section we provide a brief review of the definitions of the momentum amplituhedron $\mathcal{M}_{n,k}$ at tree level and the amplituhedron $\mathcal{A}_{n,k',L}$ at tree and loop level. This will set the stage for our new definition of the momentum amplituhedron at loop level that we introduce in the subsequent section.  

\subsection{The Momentum Amplituhedron Revisited}\label{sec:momamp}

Let us start by recalling the original definition of the (tree) momentum amplituhedron introduced in \cite{Damgaard:2019ztj}. First, we consider the Grassmannian $G(k,n)$, which is the space of all $k\times n$ matrices modulo $GL(k)$ row transformations, and the positive Grassmannian $G_+(k,n)$, which is the subset of $G(k,n)$ consisting of all positive matrices, i.e. matrices with all maximal ordered minors positive. We also introduce a pair of fixed matrices $(\Lambda,\tilde\Lambda)$ and demands that $\tilde\Lambda\in M_+(k+2,n)$ is a positive matrix and $\Lambda\in M_{+,\tau}(n-k+2,n)$ is a twisted positive matrix, i.e. a matrix whose orthogonal complement is a positive matrix. Then, the momentum amplituhedron $\mathcal{M}_{n,k}=\Phi_{\Lambda,\tilde\Lambda}(G_{+}(k,n))$, for $2\leq k\leq n-2$, is defined as the image of the positive Grassmannian $G_{+}(k,n)$ through a linear map $\Phi_{\Lambda,\tilde\Lambda}$ specified by the two fixed matrices $\Lambda$ and $\tilde\Lambda$:
\begin{equation}
\Phi_{\Lambda,\tilde\Lambda}:G_+(k,n)\to G(k,k+2)\times G(n-k,n-k+2)\,.
\end{equation}
Explicitly, we have for $(Y,\tilde Y)\in G(k,k+2)\times G(n-k,n-k+2)$ and $C=(c_{\dot\alpha i})\in G_{+}(k,n)$ that
\begin{equation}\label{eq:YYt}
\quad Y_{\alpha}^{A}=\sum_i(c^\perp)_{\alpha i}\Lambda_i^A\,,\quad\tilde Y_{\dot\alpha}^{\dot A}=\sum_ic_{\dot\alpha i}\tilde\Lambda_i^{\dot A}\,,
\end{equation}
where $(c^\perp)_{\alpha i}$  for $\alpha=1,\ldots,k$ and $i=1,\ldots, n$ are the elements of the orthogonal complement $C^\perp$ of the matrix $C$.

The linear map $\Phi_{\Lambda,\tilde\Lambda}$ can be further composed with a projection
\begin{equation}
\mathcal{P}_{\Lambda,\tilde\Lambda}:G(k,k+2)\times G(n-k,n-k+2)\to G(2,n)\times G(2,n)\,,
\end{equation}
 to extract the familiar spinor helicity variables 
\begin{equation}\label{eq:llt}
\lambda_i^a=\sum_{A}(Y^\perp)^a_A \,\Lambda_i^A\,,\quad\quad \tilde\lambda_i^{\dot{a}}=\sum_{\dot{A}}(\tilde Y^\perp)^{\dot a}_{\dot A} \,\tilde\Lambda_i^{\dot{A}}\,.
\end{equation}
The image $\overline{\mathcal{M}}_{n,k}:=(\mathcal{P}_{\Lambda,\tilde\Lambda}\circ \Phi_{\Lambda,\tilde\Lambda})(G_{+}(k,n))$ of the positive Grassmannian through the composition of these maps is then a region defined directly in spinor helicity space.
As shown in \cite{Damgaard:2019ztj}, definition \eqref{eq:YYt} implies that for any point $(\lambda,\tilde\lambda)\in \overline{\mathcal{M}}_{n,k}$ the spinor brackets $\langle ii+1\rangle:=\lambda_i^{1}\lambda_j^2-\lambda_i^{2}\lambda_j^1>0$ and $[ii+1]:=\tilde\lambda_i^1\tilde\lambda_j^2-\tilde\lambda_i^2\tilde\lambda_j^1 > 0$ are positive. Moreover, for particular choices of matrices $\Lambda$ and $\tilde\Lambda$ also all planar Mandelstam variables $s_{i,i+1,\ldots,i+j} > 0$ are positive, ensuring that the boundary structure of $\overline{\mathcal{M}}_{n,k}$ reflects the correct singularity structure of the tree-level amplitudes in planar $\mathcal{N}=4$ sYM. Additionally, the following sign flip patterns are satisfied by the spinor brackets:
\begin{align}
&\{\langle 12\rangle,\langle 13\rangle,\ldots, \langle 1n\rangle\}\mbox{ has $k-2$ sign flips}\,,\\
&\{[ 12],[ 13],\ldots, [ 1n]\}\mbox{ has $k$ sign flips}\,.
\end{align}

Before moving on to the definition of the amplituhedron, we want to make an important comment regarding the properties of $\overline{\mathcal{M}}_{n,k}$ that has not been previously spelled out in the literature but will be crucial in the following. Notice that the definitions \eqref{eq:YYt} and \eqref{eq:llt} trivially imply that for any point $(\lambda,\tilde\lambda)\in \overline{\mathcal{M}}_{n,k}$ we have
\begin{equation}
\sum_i\lambda_i^a\,(c^\perp)_{\alpha i}=0\,,\quad \sum_i \tilde\lambda_i^{\dot{a}}c_{\dot\alpha i}=0\,,
\end{equation}
and therefore\footnote{These relations between $\lambda$, $\tilde\lambda$ and $C$ were the starting point of the development of the Grassmannian approach to scattering amplitudes in \cite{Arkani-Hamed:2009ljj}. Here, instead they rather result from the definition of the momentum amplituhedron. } $\tilde\lambda\cdot C=0$ and $\lambda \subset C$. Let us make the latter statement more precise. To this extent, we will use the observation made in \cite{Damgaard:2019ztj} that allows to rewrite the $\lambda$ part of \eqref{eq:YYt} and \eqref{eq:llt} in an alternative way. Let us modify the previous definition and define a map $\phi_{\Lambda^\perp,\tilde\Lambda}$ labelled by two positive matrices $\Lambda^\perp\in M_+(k-2,n)$ and $\tilde\Lambda\in M_+(k+2,n)$:
\begin{equation}
\phi_{\Lambda^\perp,\tilde\Lambda}:G_+(k,n)\to G(2,n)\times G(2,n)\,,
\end{equation} 
that will generalize the compositions of functions in our original definition. To every element in the positive Grassmannian $G_+(k,n)$ it associates a pair of matrices $(\lambda,\tilde \lambda)$:
\begin{equation}
\lambda_{ i}^{a} =\sum_\alpha (X^\perp)_{ \alpha}^{a} c_{\alpha i}\,,\quad\quad \tilde\lambda_i^{\dot{a}}=\sum_{\dot{A}}(\tilde Y^\perp)^{\dot a}_{\dot A} \,\tilde\Lambda_i^{\dot{A}}\,,
\end{equation}
where $X_{\alpha}^{\bar{a}}=\sum_i\left(\Lambda^\perp\right)_{i}^{ \bar{a}}c_{\alpha i} \in G(k-2,k)$. Then $\overline{\mathcal{M}}_{n,k}=\phi_{\Lambda^\perp,\tilde\Lambda}(G_{+}(k,n))$. The equivalence between this definition and the one in \eqref{eq:llt} descends from the fact that for all $i,j$ we have $\langle ij \rangle=\langle Yij\rangle=(X ij)_C$, where the last equality was proven in \cite{Damgaard:2019ztj}
. Importantly, the variables $X$ that we introduced allow one to construct a $GL(k)$ matrix

\begin{equation}
G=\left(\begin{matrix}
& &\hspace{-0.9cm}{X^\perp}_{2,k}\\
\hline
0_{k-2,2} & \vline &1_{k-2,k-2}\\

\end{matrix}\right),
\end{equation}
such that 
\begin{equation}\label{eq:smallc}
GC=\left(\begin{matrix}\lambda\\c\end{matrix}\right),
\end{equation}
where $c\in M(k-2,n)$.
Therefore, for any element $C\in G_+(k,n)$ and for the corresponding image point $(\lambda(C),\tilde\lambda(C))\in\overline{\mathcal{M}}_{n,k}$, it is always possible to find a representative for $C$ such that the first two rows of the matrix $C$ are the $\lambda$s.

The final comment in this section is related to the fact that points in the positive Grassmannian space $G_+(k,n)$ featuring in the definition of the momentum amplituhedron have a well-known relation to the points in the positive Grassmannian space $G_+(k',n)=G_+(k-2,n)$ in the definition of the amplituhedron \cite{Arkani-Hamed:2009nll}. Provided $\lambda\in G(2,n)$, and using relation \eqref{eq:smallc}, one can define the matrix
\begin{equation}
\label{eq:Q}
\check{c}=Q c\,,
\end{equation}
where 
\begin{equation}\label{eq:Qij}
Q_{ij}=\frac{\langle i-1 \,i\rangle \delta_{i+1,j}+\langle i \,i+1\rangle \delta_{i-1,j}+\langle i+1 \,i-1\rangle \delta_{i,j}}{\langle i-1 \,i\rangle\langle i \,i+1\rangle}\,.
\end{equation}
Importantly, we checked in various examples that if $\lambda$ is inside the momentum amplituhedron $\overline{\mathcal{M}}_{n,k}$, then $\check{c}\in G_{+}(k-2,n)$ is a positive matrix, and we conjecture this holds for all $n$ and $k$.

\subsection{The Amplituhedron}\label{sec:amplituhedron}

Similar to the momentum amplituhedron, the $L$-loop amplituhedron $\mathcal{A}_{n,k',L}$ can be defined \cite{Arkani-Hamed:2017vfh} as the image of a particular space, generalizing the positive Grassmannian, through a linear map
\begin{equation}
\Phi_Z:G_+(k',n)\dottimes G(2,n)^{L}\rightarrow G(k',k'+4)\times G(2,k'+4)^{L}\,,
\end{equation}
where $k'=k-2$ and $Z\in M_+(k'+4,n)$ is a positive matrix. The map $\Phi_Z$ assigns to every point $C=(c_{\alpha i})\in G_{+}(k',n)$ and a collection of points $D_l=(d_{l,\gamma i})\in G(2,n)$ the values
\begin{equation}
(Y_Z)^I_\alpha=  \sum_{i=1}^n c_{\alpha i}\,Z_i^I\,,\quad \quad \mathcal{L}^{I}_{l,\gamma}=  \sum_{i=1}^n d_{l,\gamma i}\,Z_i^I\,,
\end{equation}
with $l=1,\ldots,L$ enumerating the loops and $\gamma=A,B$.
The domain $G_+(k',n)\dottimes G(2,n)^{L}$ of $\Phi_Z$ is defined as all points $(C,D_1,\ldots,D_L)\in G(k',n)\times G(2,n)\times \ldots\times G(2,n)$ such that all matrices
\begin{equation}
\left(\begin{matrix}C\end{matrix}\right) ,\quad
\left(\begin{matrix}D_{l_1}\\C\end{matrix}\right) ,\,\quad\ldots \quad
\left(\begin{matrix}D_{l_1}\\\vdots\\D_{l_L}\\C\end{matrix}\right) ,
\end{equation}
are positive for $l_i=1,\ldots,L$. Then the loop amplituhedron is defined as
\begin{equation}
\mathcal{A}_{n,k',L}=\Phi_Z(G_+(k',n)\dottimes G(2,n)^L)\,.
\end{equation}

As for the momentum amplituhedron, we can compose $\Phi_Z$ with a projection
\begin{equation}
\mathcal{P}_Z:G(k',k'+4)\times G(2,k'+4)^{L}\to G(4,n)\times G(2,4)^L\,,
\end{equation}
and define the bosonic part of the momentum twistors $z^a_i$ as 
\begin{equation}\label{eq:projectionZ}
z_i^a=\sum_{I}(Y_Z^\perp)^a_I Z_i^I\,.
\end{equation}
The momentum twistor line parametrizing the loop momenta is specified by a pair of points $(AB)_l$ where
\begin{equation}\label{eq:projectionZloop}
z_{l,\gamma}^a = \sum_{I}(Y_Z^\perp)^a_I \mathcal{L}^{I}_{l,\gamma} \,.
\end{equation}
This allows us to define the loop amplituhedron directly in the momentum twistor space
\begin{equation}
\overline{\mathcal{A}}_{n,k',L}=(\mathcal{P}_Z\circ \Phi_Z)(G_+(k',n)\dottimes G(2,n)^L)\,.
\end{equation}

For completeness, we recall an important property of $\overline{\mathcal{A}}_{n,k',L}$ that follows from the definitions of the maps $\Phi_{Z}$ and $\mathcal{P}_Z$: given a point $(z_i,(AB)_1,\ldots ,(AB)_L)\in\overline{\mathcal{A}}_{n,k',L}$, the brackets of momentum twistors $\langle i\ i+1 \, j\,j+1\rangle> 0$ are positive, where $\langle ijkl\rangle=\epsilon_{IJKL}z_i^Iz_j^Jz_k^Kz_l^L$. Similarly, for brackets involving loop variables we have $\langle (AB)_l\, i\ i+1 \rangle> 0$ and $\langle (AB)_a(AB)_b \rangle> 0$. Moreover, points in the loop amplituhedron are known to have the following sign flip patterns 
\begin{align}
\{\langle (AB)_a12\rangle,\langle (AB)_a13\rangle,\ldots,\langle (AB)_a1n\rangle\} &\mbox{ has } (k'+2) \mbox{ sign flips for each loop}\,,\nonumber \\
\{\langle 1234\rangle,\langle 1235\rangle,\ldots,\langle 123n\rangle\} &\mbox{ has } k' \mbox{ sign flips}\,.
\end{align}

\subsection{Boundaries of Amplituhedra}

The amplituhedra we reviewed in the previous sections are conjectured\footnote{There is no general proof of this fact, but in all cases where the explicit answer has been found, one can check that it is indeed true.} to define positive geometries\footnote{To be more precise they are weighted positive geometries as advocated in \cite{Dian:2022tpf}.} \cite{Arkani-Hamed:2017tmz} and therefore can be equipped with canonical differential forms that encode physical quantities.  For the tree momentum amplituhedron, the canonical form encodes the tree-level scattering amplitudes in planar $\mathcal{N}=4$ sYM, written in the non-chiral spinor helicity superspace. On the other hand, the loop amplituhedron canonical form provides the loop integrand in the same theory but written in the momentum twistor space instead. Importantly, in the latter case the tree-level MHV factor is removed. This last statement has  significant implications for the singularity structure of the canonical form of the amplituhedron, and therefore for the boundary structure of the amplituhedron itself.
In particular, at tree level, the codimension one boundaries of the momentum amplituhedron $\overline{\mathcal{M}}_{n,k}$ are given by the collinear limits, $\langle i \,i+1\rangle=0\,,  [i\,i+1]=0$, and the factorization limits $s_{i,i+1\ldots,i+p}=0\,, \textrm{with}\, p=2,\ldots,n-4$.
On the other hand, for the amplituhedron the facets are given by the points satisfying $\langle i\ i+1 \, j\,j+1\rangle = 0$. When $j=i+2$, the boundary $\langle i\, i+1\,i+2\,i+3\rangle=0$ translates into the boundary $[ ii+1] =0$ of the tree momentum amplituhedron. When $j>i+2$, the boundary corresponds to a factorization channel and one can map it to the boundary of the momentum amplituhedron where one of the planar Mandelstam variables vanishes. 
However, since the MHV amplitude is factored out for the amplituhedron, there is no boundary of the amplituhedron that would correspond to $\langle i \,i+1\rangle=0$. 
Therefore, the boundary structure of the tree momentum amplituhedron is very different from the boundary structure of the tree amplituhedron. This has the important implication that there is no simple map relating the points in the two amplituhedra.
However, there exists a direct translation of all singularities of loop-level integrands from momentum twistors  to spinor helicity space: 
\begin{eqnarray}
\langle AB i \,i+1\rangle=0 &\quad\longleftrightarrow\quad& (\ell+\sum_j p_j)^2=0\,,\\
\langle (AB)_1 (AB)_2\rangle=0 &\quad\longleftrightarrow\quad& (\ell_1 - \ell_2)^2=0\,,
\end{eqnarray}
where $\ell_l$ is a particular choice of off-shell loop momenta that we will discuss in detail in the following. Equipped with this observation, we provide in the next section the definition of the loop momentum amplituhedron.


\section{The Loop Momentum Amplituhedron}
\label{sec:loopmomamp}

We want to extend the definition of momentum amplituhedron from the previous section to include loops. As we argued before, since there is a one-to-one correspondence between singularities of loop-level integrands written  in terms of momentum twistors and in the momentum space (after we translate between kinematic variables), we will adapt the construction of the loop amplituhedron from section \ref{sec:amplituhedron} into spinor helicity variables. To start, let us recall that in the momentum twistor space each loop momentum is encoded as a line $(AB)_l$ that can be defined by specifying two momentum twistors $z_{l,A}$ and $z_{l,B}$, up to a $GL(2)$ transformation. To simplify our notation, we will drop the loop index $l$ in the remaining part of this section, and discuss a single loop variable. In the amplituhedron construction, after we use the projections \eqref{eq:projectionZ} and \eqref{eq:projectionZloop}, the line is parametrised in terms of the external momentum twistors as
\begin{equation}\label{eq:ZABinD}
z_{\gamma}^a=\sum_{i}d_{\gamma i}z_i^a\,,\qquad \gamma=A,B\,.
\end{equation}
Moreover, momentum twistors can be written in terms of the spinor helicity variables $\lambda$ and dual space coordinates $x_i$ as
\begin{equation}
z_i=\left(\begin{matrix}\lambda_i\\\tilde\mu_i\end{matrix}\right)=\left(\begin{matrix}\lambda_i\\x_i \lambda_i\end{matrix}\right)\,,\qquad i=1,\ldots ,n\,,
\end{equation}
and similarly for the loop momentum twistors in terms of a single dual space coordinate $x$:
\begin{equation}\label{eq:ZABinx}
z_{\gamma}=\left(\begin{matrix}\lambda_{\gamma}\\\tilde\mu_{\gamma}\end{matrix}\right)=\left(\begin{matrix}\lambda_{\gamma}\\x \lambda_{\gamma}\end{matrix}\right)\,,\qquad \gamma=A,B\,.
\end{equation}

Combining the expansion \eqref{eq:ZABinD} with \eqref{eq:ZABinx}, we immediately get an expansion of $\lambda_{A}$ and $\lambda_{B}$ in terms of the external particles:
\begin{equation}
\lambda_{\gamma}^\alpha=\sum_{i}d_{\gamma i}\lambda_i^\alpha\,,\qquad \gamma=A,B\,.
\end{equation} 
Since $D=(d_{\gamma i})\in G(2,n)$, then there is a natural $GL(2)$ transformation between $\lambda_{A}$ and $\lambda_{B}$. 

One of the most important insights from this simple calculation is that, if we want to translate the momentum twistors loop variables to spinor helicity space, we should look for a parametrisation of the loop momenta that renders manifest this $GL(2)$ transformation. In the following, we will use the following parametrisation of off-shell momentum, written in terms of spinor helicity variables
\begin{equation}
\label{loop_def}
\ell=\lambda_A \tilde\lambda^A+\lambda_B \tilde \lambda^B\,,
\end{equation}
where, in order for $\ell$ to be $GL(2)$-invariant, $\tilde\lambda^A$, $\tilde\lambda^B$ need to transform as
\begin{equation}
\left(\begin{matrix}\lambda'_A&\lambda'_B\end{matrix}\right)=\left(\begin{matrix}\lambda_A&\lambda_B\end{matrix}\right)\cdot G\,, \qquad\qquad \left(\begin{matrix}\tilde\lambda'^A\\\tilde\lambda'^B\end{matrix}\right)=G^{-1}\cdot\left(\begin{matrix}\tilde\lambda^A\\\tilde\lambda^B\end{matrix}\right),
\end{equation}
for $G\in GL(2)$. 

In the next step, we want to derive the expansion of $\tilde\lambda^{A}$ and $\tilde\lambda^{B}$ in terms of external particles by considering the remaining part of the condition \eqref{eq:ZABinD}. First, let us introduce
\begin{equation}
\tilde\lambda_\gamma=\sum_{\delta=A,B}\epsilon_{\gamma\delta}\tilde\lambda^\delta \,,\quad \gamma=A, B \,.
\end{equation}
Taking the last two entries in the expansion \eqref{eq:ZABinD}, we have that
\begin{equation}
\tilde\mu_\gamma=\sum_i d_{\gamma i}\tilde\mu_i  
\end{equation} 
with $\tilde\mu_i=x_i \lambda_i$ and $\tilde\mu_\gamma=x_\gamma\lambda_\gamma$. A simple calculation results in
\begin{align}\label{eq:xl-x1l}
x \lambda_A=\sum_{i}d_{Ai}x_i \lambda_i=\sum_{i}d_{Ai}\left(x_1-\sum_{j=1}^{i-1}p_j\right) \lambda_i=x_1 \lambda_A-\sum_{j<i}d_{Ai}\langle ji\rangle \tilde\lambda_j\,.
\end{align}
As the last step, we need to identify the loop momentum $\ell$ with the dual coordinate $x$. There are various choices that we could make which would result in different loop momentum amplituhedron geometries. Importantly, the canonical forms on these geometries can be related to each other by the change of variables between these choices. Motivated by the explicit form of the relation \eqref{eq:xl-x1l}, we settled for the following relation
\begin{equation}
\label{loop_def2}
\ell=x-x_1=\lambda_A \tilde\lambda^A+\lambda_B \tilde \lambda^B=\lambda_A\tilde\lambda_B-\lambda_B\tilde\lambda_A \,.
\end{equation} 
This allows us to find the explicit expansion of $\tilde\lambda_\gamma$ in terms of the external particles
\begin{equation}
\tilde\lambda_\gamma=\sum_{j<i}d_{\gamma i}\frac{\langle ij\rangle}{\langle AB\rangle}\tilde\lambda_j\,.
\end{equation}
Using \eqref{loop_def2}, we find the loop momentum written in terms of external spinor helicity variables, as well as of elements of the matrix $D$:
\begin{equation}
\ell=\left(\sum_i d_{Ai}\lambda_i\right)\left(\sum_{j<i}d_{Bi}\frac{\langle ij\rangle}{\langle AB\rangle}\tilde\lambda_j\right)-\left(\sum_i d_{Bi}\lambda_i\right)\left(\sum_{j<i}d_{Ai}\frac{\langle ij\rangle}{\langle AB\rangle}\tilde\lambda_j\right).
\end{equation}
Importantly, this formula provides a \emph{global} definition of loop momentum. 

We are now ready to define the loop momentum amplituhedron. To this effect, we extend the map from section \ref{sec:momamp} to include also loop momenta. We define 
\begin{align*}
\begin{matrix}
\tilde\phi_{(\Lambda^\perp,\tilde\Lambda)}:&G_+(k,n)&\dottimes &G(2,n)^{L}&\to &G(2,n)&\times &G(2,n) &\times &GL(2)^{{L}}\\
&C&&D_l&\mapsto&\lambda&&\tilde\lambda&&\ell_l
\end{matrix}
\end{align*}
where $\Lambda^\perp\in M_+(k-2,n)$, $\tilde\Lambda\in M_+(k+2,n)$ and we will define the product $\dottimes$ shortly. The map $\tilde\phi_{(\Lambda^\perp,\tilde\Lambda)}$ associates to every point $C\in G_+(k,n)$ and a collection of points $D_l\in G(2,n)$, the tree-level variables $(\lambda, \tilde\lambda)$ given by \eqref{eq:llt} and the loop momenta $\ell_l$ given by \eqref{loop_def2}.
To complete our definition, we need to explain what we mean by the $\dottimes$ product present in the domain of  $\tilde\phi_{(\Lambda^\perp,\tilde\Lambda)}$. We have already introduced in \eqref{eq:Qij} the matrix $Q(\lambda)$ that relates the Grassmannian points in the definition of the tree amplituhedron to the Grassmannian points in the definition of the tree momentum amplituhedron. In particular, we conjectured that for $(\lambda,\tilde\lambda)\in \overline{\mathcal{M}}_{n,k}$ and $C\in G_+(k,n)$, if we can define $\check{c}=Q\cdot c$ then $\check{c}\in G_{+}(k-2,n)$. The product $\dottimes$ is defined by additional positivity conditions relating the matrix $\check{c}$ with the loop-level matrices $D_l$. We define  $G_+(k,n)\dottimes G(2,n)^{L}$ as the set of all points $C\in G_+(k,n)$ and $D_l\in G(2,n)$ for $l=1,\ldots L$ such that all matrices
\begin{equation}\label{eq:posccheck}
\left(\begin{matrix}\check{c}\end{matrix}\right) ,\quad
\left(\begin{matrix}D_{l_1}\\\check{c}\end{matrix}\right) ,\,\quad\ldots \quad
\left(\begin{matrix}D_{l_1}\\\vdots\\D_{l_L}\\\check{c}\end{matrix}\right) ,
\end{equation}
are positive for all $l_i=1,\ldots,L$. 

The {\it loop momentum amplituhedron} $\overline{\mathcal{M}}_{n,k,L}$ is then defined as the image
\begin{equation}
\overline{\mathcal{M}}_{n,k,L}=\tilde\phi_{(\Lambda^\perp,\tilde\Lambda)}\left(G_+(k,n)\dottimes G(2,n)^{L}\right)\,.
\end{equation}
This is the main result of our paper.

We conclude this section by having a preliminary look at the boundary structure of the loop momentum amplituhedron. First, it is clear from our construction that $\overline{\mathcal{M}}_{n,k,L}$ has boundaries when the tree-level invariants vanish, reflecting the facet structure of the tree momentum amplituhedron. In particular, it has boundaries that correspond to factorisations when the planar Mandelstam variables vanish $s_{i,i+1,\ldots,i+p}= 0$. It also has boundaries coming from collinear limits of two types\footnote{It is possible that these boundaries are not facets of the loop momentum amplituhedron but instead they are lower dimensional, as for example in the 4-point case.} when $\langle ii+1\rangle= 0$ or $[ii+1]= 0$. These are supplemented by the codimension-one boundaries of two types coming from loop level: 
\begin{itemize}
\item $(\ell_r+\sum_j p_j)^2=0$ corresponding to a sufficient number of minors of the matrix $\left(\begin{matrix}D_r\\\check{c}\end{matrix}\right)$ vanishing for some $r=1,\ldots,L$;
\item $(\ell_{r_1}-\ell_{r_2})^2=0$ corresponding to a sufficient number of minors of the matrix $\left(\begin{matrix}D_{r_1}\\D_{r_2}\\\check{c}\end{matrix}\right)$ vanishing for some $r_1\neq r_2$.
\end{itemize}
Finding the complete stratification of boundaries of $\overline{\mathcal{M}}_{n,k,L}$ remains an open and interesting problem that we plan to address in the future.


\section{Examples}
\label{sec:examples}

In this section we present a few examples of the loop momentum amplituhedron and the related amplitudes.

Let us start with the simplest case, the MHV amplitudes. Since in this case we have $k'=k-2=0$, then the matrix $\check c$ in \eqref{eq:posccheck} is an empty matrix and there is no positivity condition mixing $C$ and the loop matrices $D_l$. This means that each matrix $D_l\in G_+(2,n)$ is positive on its own, and there are additional mutual positivity conditions between $D$s as in \eqref{eq:posccheck}. Therefore, the geometry in this case is simply the product of the tree-level geometry and the loop geometry. This immediately implies that the canonical form $\Omega_{n,2,L}$ for the loop momentum amplituhedron $\mathcal{M}_{n,2,L}$ is the wedge product of the canonical form for the tree momentum amplituhedron $\mathcal{M}_{n,2,0}$ times the $4L$-form coming from the loop geometry:
\begin{equation}
\Omega_{n,2,L}=\Omega_{n,2,0} \wedge  \tilde\Omega_{n,2,L}\,,
\end{equation}
where $\tilde\Omega_{n,2,L}$ is a $4L$-form coming purely from the loop geometry.
Since we know from \cite{Damgaard:2019ztj} the exact form of the tree-level canonical form, we just need to find the canonical form of the loop geometry.

Because of the factorisation of the tree and loop geometries we described above, it is possible for a direct translation of the loop canonical forms from the loop amplituhedron \cite{Arkani-Hamed:2013jha}. In particular,  the one-loop amplituhedron $\overline{\mathcal{A}}_{n,0,1}$ is the union of images of the so-called kermits \cite{Arkani-Hamed:2010zjl,Arkani-Hamed:2013jha} with associated matrices:
\begin{equation}
K_{i,j} :
\begin{pmatrix}
1 &0  &\ldots & *_i & *  &\ldots&0& 0&\ldots\\
1 &0  &\ldots & 0&0&  \ldots& *_j& *&\ldots
\end{pmatrix},
\end{equation}
through the map $\mathcal{P}_Z\circ \Phi_Z$.  
Then,  we conjecture that  also the one-loop momentum amplituhedron $\overline{\mathcal{M}}_{n,2,1}$ is the union of the images of the kermits
\begin{equation}
\overline{\mathcal{M}}_{n,2,1}=\bigcup_{i<j} \tilde\phi_{(\Lambda^\perp,\tilde\Lambda)} (C, K_{i,j})\,.
\end{equation}
The canonical form $\omega_{n,0,1}$ of $\overline{\mathcal{A}}_{n,0,1}$ is known and reads
\begin{equation}
\omega_{n,0,1}=\sum_{1<i<j<n} \omega_{K_{i,j}}\,,
\end{equation}
where
\begin{equation}
\omega_{K_{i,j}} = \mathrm{dlog} \frac{\langle AB  1i \rangle}{\langle AB  1 i+1 \rangle} \wedge \mathrm{dlog} \frac{\langle AB  i i+1 \rangle}{\langle AB  1 i+1 \rangle} \wedge\mathrm{dlog} \frac{\langle AB  j j+1 \rangle}{\langle AB  1 j \rangle} \wedge\mathrm{dlog} \frac{\langle AB  j+1  1 \rangle}{\langle AB  1 j \rangle} \,.
\end{equation}
We claim that the canonical form for the one-loop momentum amplituhedron is
\begin{equation}
\Omega_{n,2,1}=\Omega_{n,2,0} \wedge \sum_{i<j} \Omega_{K_{i,j}}\,,
\end{equation}
where
\begin{equation}
\Omega_{K_{i,j}}=\mathrm{dlog\frac{(\ell-\ell^*_{1\,i})^2}{(\ell-\ell^*_{1\,i+1})^2}}\wedge
\mathrm{dlog\frac{(\ell-\sum_{a=1}^ip_a)^2}{(\ell-\ell^*_{1\,i+1})^2}}\wedge \mathrm{dlog\frac{(\ell-\ell^*_{1\,j})^2}{(\ell-\ell^*_{1\,j+1})^2}} \wedge \mathrm{dlog\frac{(\ell-\sum_{a=1}^jp_a)^2}{(\ell-\ell^*_{1\,j+1})^2}}\,,
\end{equation}
and we have defined
\begin{equation}
\ell_{ij}^*=\frac{1}{\langle ij \rangle }\left(\lambda_i \sum_{l=1}^{j-1}\langle lj\rangle\tilde\lambda_l - \lambda_j \sum_{l=1}^{i-1}\langle li\rangle\tilde\lambda_l \right).
\end{equation}

\begin{figure}[t]

\begin{center}
\includegraphics[scale=0.18]{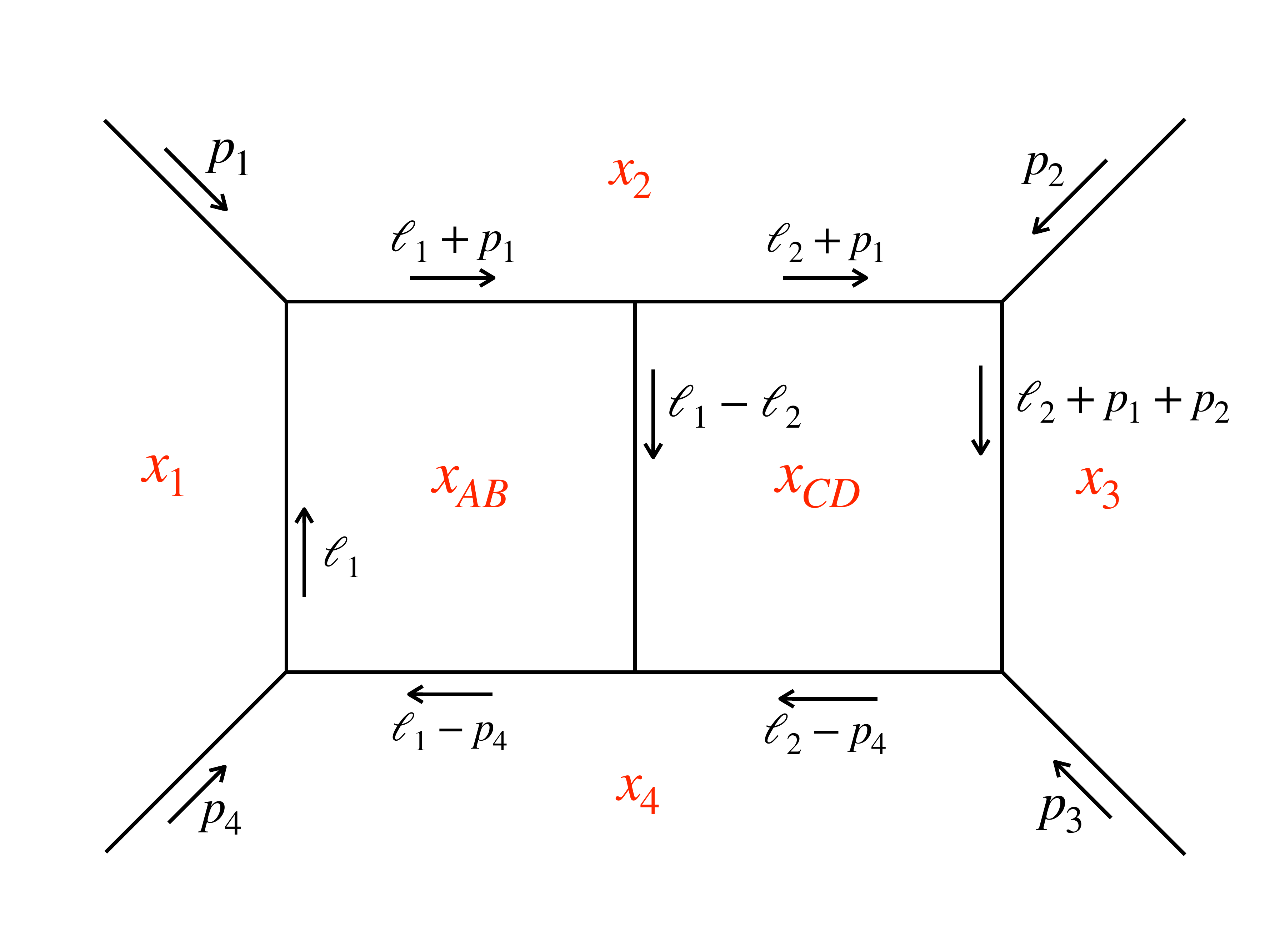}
\includegraphics[scale=0.24]{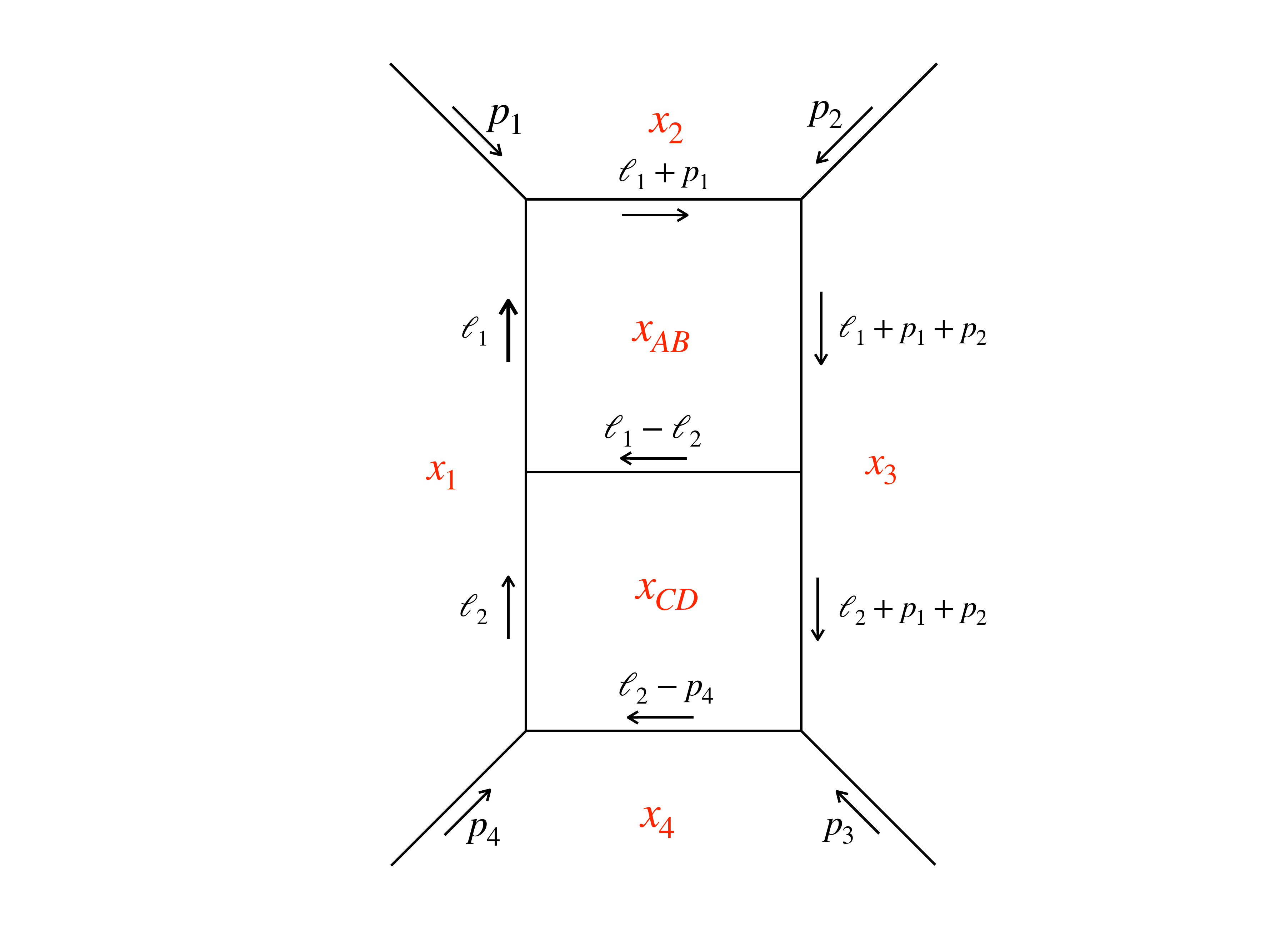}
\label{fig:2loop}
  \caption{The two diagrams corresponding to the two terms in \eqref{twoloop}. The remaining two terms have identical diagrams with $x_{AB}$ and $x_{CD}$ exchanged.}
  \end{center}
\end{figure}

Moving on beyond one loop, general triangulations of the loop amplituhedron can be obtained from the BCFW recursion relation \cite{Arkani-Hamed:2010zjl} together with the on-shell diagram parametrisation proposed in \cite{Bai:2014cna}. We conjecture that for MHV amplitudes, the images of the same BCFW matrices will triangulate the loop momentum amplituhedron and loop amplituhedron. To support our claim, we provide the simplest example beyond one loop: two-loop four-point amplitude. In this case, there are 16 BCFW terms \cite{Bai:2014cna} and we have performed extensive numerical checks that given a set of positive data and a point inside the loop momentum amplituhedron, it lies in one and only one of the images of the 16 cells corresponding to these BCFW terms. Then we can translate the known canonical forms to spinor helicity space and sum them together. Ultimately, we get the well-known formula
\begin{eqnarray}
\label{twoloop}
\hspace{-0.5cm}\tilde\Omega_{4,2,2} &=& \left\{\frac{s^2 t \, \mathrm{d}^4\ell_1 \mathrm{d}^4\ell_2}{\ell_1^2 (\ell_1+p_1)^2 (\ell_1-p_4)^2 (\ell_1-\ell_2)^2 (\ell_2+p_1)^2 (\ell_2+p_1+p_2)^2 (\ell_2-p_4)^2} \right. \nonumber \\
&+&\left. \frac{s t^2 \, \mathrm{d}^4\ell_1 \mathrm{d}^4\ell_2}{\ell_1^2 (\ell_1+p_1)^2 (\ell_1+p_1+p_2)^2 (\ell_1-\ell_2)^2  \ell_2^2  (\ell_2+p_1+p_2)^2 (\ell_2-p_4)^2}\right\} + (\ell_1 \leftrightarrow \ell_2)\,,
\end{eqnarray} 
where as defined before we have $\ell_1=x_{AB}-x_1$ and $\ell_2=x_{CD}-x_1$. Each term in this expansion corresponds to the expression associated to a  standard Feynman diagram, see  fig. \ref{fig:2loop}.
Since for MHV amplitudes the parametrization of the BCFW cells is known at any loop \cite{Bai:2014cna}, then it is in principle possible to extend our calculation beyond four points and beyond two loops to find canonical differential forms for the loop momentum amplituhedron $\overline{\mathcal{M}}_{n,2,L}$.

We conclude this section by having a first look at examples beyond the MHV sector. In this case, the geometry is not the product of the tree-level and loop geometries anymore. Indeed, the matrix $\check c$ is not empty, and therefore there are positivity conditions mixing the matrices $\check c$ and $D_l$.
For instance, for next-to-MHV (NMHV) amplitudes at one loop, we require that the $3 \times n$ matrix $C$ is positive (that implies that the $1 \times n$ matrix $\check c$ is positive), and the  $3 \times n$ matrix obtained by stacking a single $D$ matrix on top of the $\check c$ is positive. Let us consider the 5-point case. One would naively think to be able to use the BCFW triangulations of the loop amplituhedron given in \cite{Arkani-Hamed:2013jha,Bai:2015qoa}, where one can find three BCFW terms with their parametrisations of $(D,\check{c})$ matrices. Then, one considers points in the domain of the loop momentum amplituhedron map $C\dottimes D$ corresponding to points in these BCFW cells, namely points for which $\check{c}=Q c$. Since $Q$ has rank $n-2$ and is therefore not an invertible matrix, this is however not possible\footnote{It differs at tree level where one can construct a map from positroid cells of $G_+(k-2,n)$ to positroid cells $G_{k,n}$, the so-called T-duality map \cite{Lukowski:2020dpn,Arkani-Hamed:2012zlh}. The T-duality map however acts on whole positroid cells and not on their points, as is required at loop level.}. Therefore, it is a non-trivial task to find triangulations of the loop momentum amplituhedron using the known results about the loop amplituhedron, and the problem of triangulating $\overline{\mathcal{M}}_{n,k,L}$ for $k>2$ remains the most urgent unresolved question. The fact that the triangulations of amplituhedron and momentum amplituhedron cannot be easily matched also means that the geometry of the loop momentum amplituhedron beyond MHV level is much richer in structure and deserves further study.


\section{Conclusions and Outlook}

In this paper we presented the geometry for scattering amplitudes in $\mathcal{N}=4$ sYM at tree and loop level  in spinor helicity space, i.e. the loop momentum amplituhedron.  Taking inspiration from the singularity structure of amplitudes and expectation values of Wilson loops, we used the known construction of the loop amplituhedron and adapted it to spinor helicity space. 
Importantly, while all facets of the loop part of the amplituhedron are mapped to facets of the momentum amplituhedron at loop level, the complete boundary stratification of the two geometries is different, due to the differences at tree level. 

There are many natural questions which arise from this work. The most pressing direction is to investigate how to triangulate the loop momentum amplituhedron geometry. Unlike for the MHV loop momentum amplituhedron, where the triangulation can be directly obtained from the triangulations of the loop amplituhedron, for higher helicity sectors it is not possible anymore due to the mixing of tree and loop geometries. As for the momentum twistor space, the most natural starting point would be the BCFW recursion relation solved in terms of on-shell diagrams, which should provide parametrisations for the tree-level matrix $C$ and the loop-level matrices $D_p$.

An equally pressing question is the boundary structure of the loop momentum amplituhedron. The full stratification at tree level was found in \cite{Ferro:2020lgp} and it possesses very natural physical properties, with all boundaries labelled by Grassmannian forests that physically correspond to all possible factorisations and soft and collinear limits of tree amplitudes. We expect that also at loop level one will be able to introduce a natural, physically motivated labelling for all boundaries of the loop momentum amplituhedron. As the starting point, one can expand the methods implemented in the Mathematica package \texttt{amplituhedronBoundaries} \cite{Lukowski:2020bya} that have been crucial at tree level.  
This would provide us with a classification of all singularities of amplitude integrands at any loop order.

Another interesting question is the extension of our construction to the loop level of the orthogonal momentum amplituhedron, i.e. the positive geometry for tree-level amplitudes in ABJM theory \cite{Huang:2021jlh,He:2021llb},  that is defined in terms of the positive orthogonal Grassmannian \cite{Huang:2013owa}. 
For four-point ABJM amplitudes a geometry encoding all-loop amplitude integrands has already been suggested in \cite{He:2022sas}. This has been done by projecting the $\mathcal{N}=4$ sYM loop amplituhedron to three dimensions. Then, a natural question is if this result, combined with our definition of the loop momentum amplituhedron, allows for generating the all-loop orthogonal momentum amplituhedron also for other multiplicities.

Furthermore, it has been recently showed that the so-called ``negative geometries'' \cite{Arkani-Hamed:2021iya} provide a geometric definition of an infrared finite quantity interpreted as the expectation value of the Wilson loop with a single Lagrangian insertion, at least for four points. Following the same logic, it would be interesting to study whether one can retrieve infrared finite information about integrated amplitudes also from the loop momentum amplituhedron.

Finally, since the loop momentum amplituhedron is defined in spinor helicity space, it should allow for generalisations to the non-planar sector of $\mathcal{N}=4$ sYM, which would result in a geometry for non-planar loop integrands. 
A strong suggestion that such geometry should exist come from the fact that also the non-planar loop amplitude integrands have logarithmic singularities and can be converted to logarithmic differential forms  \cite{Arkani-Hamed:2014via}. One of the main difficulties of moving to non-planar amplitudes had been the absence of a global definition of the loop momentum. However, since our construction provides such a global definition by defining the loop momentum as a parameter of the map $\tilde\phi_{(\Lambda^\perp,\tilde\Lambda)}$, a natural conjecture would be to find non-planar contributions by modifying the domain of the map $\tilde\phi$.
This conjecture is reinforced by the fact that non-planar on-shell diagrams \cite{Arkani-Hamed:2014bca,Bourjaily:2016mnp,Paranjape:2022ymg}, which represent cuts of non-planar loop amplitudes, are connected to parts of Grassmannian spaces different from the positive one. Therefore, they can provide a useful hint on finding a non-planar momentum amplituhedron.


\section{Acknowledgements}

We would like to thank A. Lipstein and J. Trnka for useful discussions. 
This work was partially funded by the Deutsche Forschungsgemeinschaft (DFG, German Research Foundation) -- Projektnummern 404358295 and 404362017. This research was partially supported by the Munich Institute for Astro-, Particle and BioPhysics (MIAPbP) which is funded by the Deutsche Forschungsgemeinschaft (DFG, German Research Foundation) under Germany's Excellence Strategy -- EXC-2094 -- 390783311.

\appendix
\section{Kinematic Variables}
\label{app:vars}

In this appendix we  collect the variables used in $\mathcal{N}=4$ sYM which are mentioned in the paper.


\paragraph{Spinor helicity variables.}

In a massless theory  in four dimensions with $p_i^2=0$ for all particles, one can write each momentum as
\begin{equation}
p_i^{a\dot a}=\lambda_i^a\widetilde\lambda_i^{\dot a}\,,
\end{equation} 
in terms of two spinor variables $\lambda$ and $\widetilde\lambda$.
In $\mathcal{N}=4$ SYM, we can consider 
\begin{itemize}
\item the chiral superspace $(\lambda^{\alpha},\tilde\lambda^{\dot\alpha}|\eta^A)$:   $\eta^A$ are Grassmann-odd variables transforming in the fundamental representation of the $SU(4)$ R-symmetry, 
\item the non-chiral superspace $(\lambda^{\alpha},\eta^a | \tilde\lambda^{\dot\alpha},\tilde\eta^{\dot a})$: $\eta^a,\widetilde\eta^{\dot a}$ are two sets of Grassmann-odd variables both transforming in the fundamental representations of $SU(2)$. One can think of $\widetilde \eta^{\dot a}$ as Fourier conjugate variables to $\eta^{3,4}$. 
\end{itemize}

 
\paragraph{Dual superspace.}

Starting from the on-shell chiral superspace, one can define a dual superspace with coordinates $(x,\theta)$ with
\begin{equation}
x^{a\dot a}_i - x^{a\dot a}_{i-1} = \lambda^a_{i-1} \widetilde \lambda^{\dot a}_{i-1}\,,\qquad \theta^{a A}_i - \theta^{aA}_{i-1} = \lambda^a_{i-1} \eta^{ A}_{i-1}\, \quad  i = 1, \ldots , n\,.
\end{equation}
This is the space where the $n$-sided null polygon Wilson loop dual to the $n$-point amplitude is  formulated.

 
\paragraph{Momentum twistor variables.}	

The (super) momentum twistors are defined from the dual superspace
\begin{equation}
\label{eq:momtw}
\mathcal{Z}_i = (z^a_i|\chi^A_i) = (\lambda_{i a}, \tilde\mu_i^{\dot a}|\chi^A_i) \equiv (\lambda_{i a}, x^{a\dot a}\lambda_{ia}|\theta_i^{aA}\lambda_{ia})\,.
\end{equation}
The momentum twistors are unconstrained and they determine $\widetilde\lambda,\eta$ via,
\begin{equation}
(\widetilde\lambda|\eta)_i =\frac{\langle i-1\,i\rangle (\tilde\mu|\chi)_{i+1}+\langle i+1\,i-1\rangle (\tilde\mu|\chi)_{i}+\langle i\,i+1\rangle (\tilde\mu|\chi)_{i-1}}{\langle i-1\,i\rangle\langle i\,i+1\rangle}\,,
\end{equation}
and we also have
\begin{equation}
x_{ij}^2 := (x_i-x_j)^2=\frac{\langle i-1\, i\, k-1\, k\rangle}{\langle i-1\, i\rangle \langle k-1\, k\rangle} \,,
\end{equation}
where $\langle ijkl\rangle=\epsilon_{IJKL}z_i^Iz_j^Jz_k^Kz_l^L$.
Finally, in momentum twistor variables, the loop integral is an integral over the space of lines $(AB)$. This can be rewritten as an integral over a pair of points $A$ and $B$, modulo the $GL(2)$ redundancies labeling their positions on the line:
\begin{equation}
d^4\ell = \frac{d^4z_{A}d^4z_{B}}{\mathrm{vol(GL(2))}} \,.
\end{equation}

\bibliographystyle{nb}

\bibliography{momampl_loop}

\end{document}